\documentclass[aps,prl,showkeys,amsmath,amssymb,twocolumn]{revtex4}
\usepackage{graphicx}
\usepackage{bm}

\begin{document}
\title{A note on the best invariant estimation of continuous probability distributions \\under mean square loss}
\author{Thomas Sch\"urmann}
\email{t.schurmann@icloud.com}
\affiliation{J\"ulich Supercomputing Centre, J\"ulich Research Centre, 52425 J\"ulich, Germany}

\begin{abstract}
We consider the nonparametric estimation problem of continuous probability distribution functions. For the
integrated mean square error we provide the statistic corresponding to the
best invariant estimator proposed by Aggarwal \cite{Ag55} and Ferguson \cite{Fe67}. The table of critical values is
computed and a numerical power comparison of the statistic with the traditional
Cram\'{e}r-von Mises statistic is done for several representative distributions.
\end{abstract}

\maketitle
\vspace{0.5cm}
In 1933, Kolmogogov \cite{Ko33} formally defined the empirical distribution function $F_n(x)$, and
then considered how close this would be to the true distribution function $F(x)$ when it
is continuous. This contribution introduced the use of $F_n(x)$ as an estimator
of $F(x)$, to be followed by its use in testing a given $F(x)$.
Modifying the statistics proposed earlier by Cram\'{e}r \cite{Cr28} and
von Mises \cite{Mi31}, Smirnov \cite{Sm36}\cite{Sm37} compared the hypothesis $F(x)$ with $F_n(x)$ by means of the
quadratic loss function
\begin{eqnarray}\label{omega1}
L = \int_{-\infty}^\infty\,(F(x)-F_n(x))^2\,w(F(x))\,dF(x)
\end{eqnarray}
where $w$ is some preassigned positive weight function.
Let $x_1,...,x_n$ be a random sample drawn from the continuous
probability distribution function $F(x)$ with density function $f(x)$ and let
$x_1^*< x_2^*<...< x_n^*$
be obtained by ordering each realization $x_1,...,x_n$. For $w=1$, the expression (\ref{omega1}) can be
written equivalently as
\begin{eqnarray}\label{omega2}
\omega^2 = \frac{1}{12n^2}+\frac{1}{n}\sum\limits_{i=1}^n \Big[\,F_i-\frac{2i-1}{2n}\,\Big]^2
\end{eqnarray}
with $F_i=F(x_i^*)$. The term $n \omega^2$ is commonly named as the Cram\'{e}r-von\,Mises statistic. Smirnov \cite{Sm36}\cite{Sm37}\cite{Sm39} showed that the probability distribution of the latter is independent of $F$ for any $n$ and he obtained an asymptotic expression of its probability distribution for $n\to\infty$. For general weight functions, Anderson and Darling \cite{AD52}\cite{AD54} presented a general method for obtaining the asymptotic distribution of (\ref{omega1}) for $n\to\infty$.

Later on, Aggarwal \cite{Ag55} considered a class of invariant loss functions and
obtained the best invariant estimators which are also step functions like $F_n(x)$.
The canonical representation of any invariant estimator is given by
\begin{eqnarray}\label{Fe}
\hat{F}(x)=\sum_{i=0}^n \hat{F}_i\,\textbf{1}_{\nu_i}(x)
\end{eqnarray}
with real valued factors $0\leq\hat{F}_i\leq 1$, $i=0,...,n$, and $\textbf{1}_{\nu_i}(x)$ is the indicator function of the set
$\nu_i$, which is defined by
\begin{eqnarray}\label{part1}
\nu_0 &=& (\,-\infty ,\, x_1^*\,)\nonumber\\
\nu_i &=& [\, x_i^* ,\, x_{i+1}^*)\qquad i=1,...,n-1,\\
\nu_n &=& [\,x_n^*,\infty\,).\nonumber
\end{eqnarray}
Note that for the empirical distribution, each set $\nu_i$ is
uniquely corresponding to the particular probability estimate $i/n$, for $i=0,...,n$.\\
For the risk function
\begin{eqnarray}\label{R}
R= \text{E}\!\int\,(F-\hat{F})^2\,w(F)\,dF
\end{eqnarray}
it is known \cite{Ag55}\cite{Fe67} that
\begin{eqnarray}\label{F1}
\hat{F}(x)=F_n(x)
\end{eqnarray}
is the best invariant estimate if the weight function is $w(t)=1/t(1-t)$, while the
corresponding statistic (\ref{omega1}) is given by
\begin{eqnarray}\label{A}
A^2=-1-\frac{1}{n}\sum_{i=1}^n \frac{2i-1}{n}\Big[\,\log F_i+\log(1-F_{n-i+1})\,\Big].
\end{eqnarray}
The expression $nA^2$ is the commonly denoted Anderson-Darling statistic. On the other hand,
it is also known that the estimator
\begin{eqnarray}\label{F2}
\hat{F}(x)=\frac{n F_n(x)+1}{n+2}
\end{eqnarray}
is best invariant for the ordinary mean square error with $w(t)=1$ in (\ref{omega1}). Although the latter can be considered as an improvement of the Cram\'{e}r-von Mises statistic, it appears that there is no explicit expression of the corresponding statistic given in literature.

Read \cite{Re72} pointed out that (\ref{F1}) and (\ref{F2}) can be improved by an estimator which is stochastically smaller than the Kolmogorov statistic. However, the corresponding estimator is not a step function and is not invariant under the full group of strictly increasing transformations as are (\ref{F1}) and (\ref{F2}). Another discussion concerns the \textit{admissibility} of the estimators (\ref{F1}) and (\ref{F2}). For instance, Yu \cite{Yu89a}\cite{Yu89b}\cite{Yu89c} has shown that (\ref{F1}) is admissible with respect to the weight $w(t)=1/t(1-t)$ only for samples of size $n=1,2$ but inadmissible for samples of size $n\geq 3$. Similarly, for the estimator (\ref{F2}), Brown \cite{Br88} has proven inadmissibility for all sample sizes $n\geq 1$. Nevertheless, it is not known if the alternative estimator in Brown's proof is by itself admissible or whether it is possible to find some other estimator dominating (\ref{F2}) which provides a significantly larger saving in risk from using (\ref{F2}). Therefore, let us provide the following:\\
\\
\textbf{Proposition 1.} For unit weight $w=1$, the statistic (\ref{omega1}) corresponding to the best invariant estimator (\ref{F2}) is
\begin{eqnarray}\label{SH}
\hat\omega^2=\frac{n+8}{12(n+2)^3}+\frac{1}{n+2}\,\sum_{i=1}^n\,\Big[\,F_i-\frac{\,i+\frac{1}{2}}{\,n+2\,}\,\Big]^2.
\end{eqnarray}
\textbf{Proof.} To obtain (\ref{SH}) one has to replace $F_n(x)$ in (\ref{omega1}) by the estimator (\ref{F2}) and compute the integration. After some algebraic manipulations expression (\ref{SH}) is obtained.\hfill$\Box$\\
\\
The statistic $\hat\omega^2$ is similar but different from the traditional Cram\'{e}r-von Mises
statistic (\ref{omega2}). For large sample sizes they have similar stochastic properties. The minimum risk
of the statistic is $1/6(n+2)$, which is slightly less than for the Cram\'{e}r-von\,Mises statistic with $1/6n$.
Thus, at least for small samples we can expect that (\ref{SH}) will improve (\ref{omega2}).

\begin{figure*}[thb]
\includegraphics[width=12.0cm]{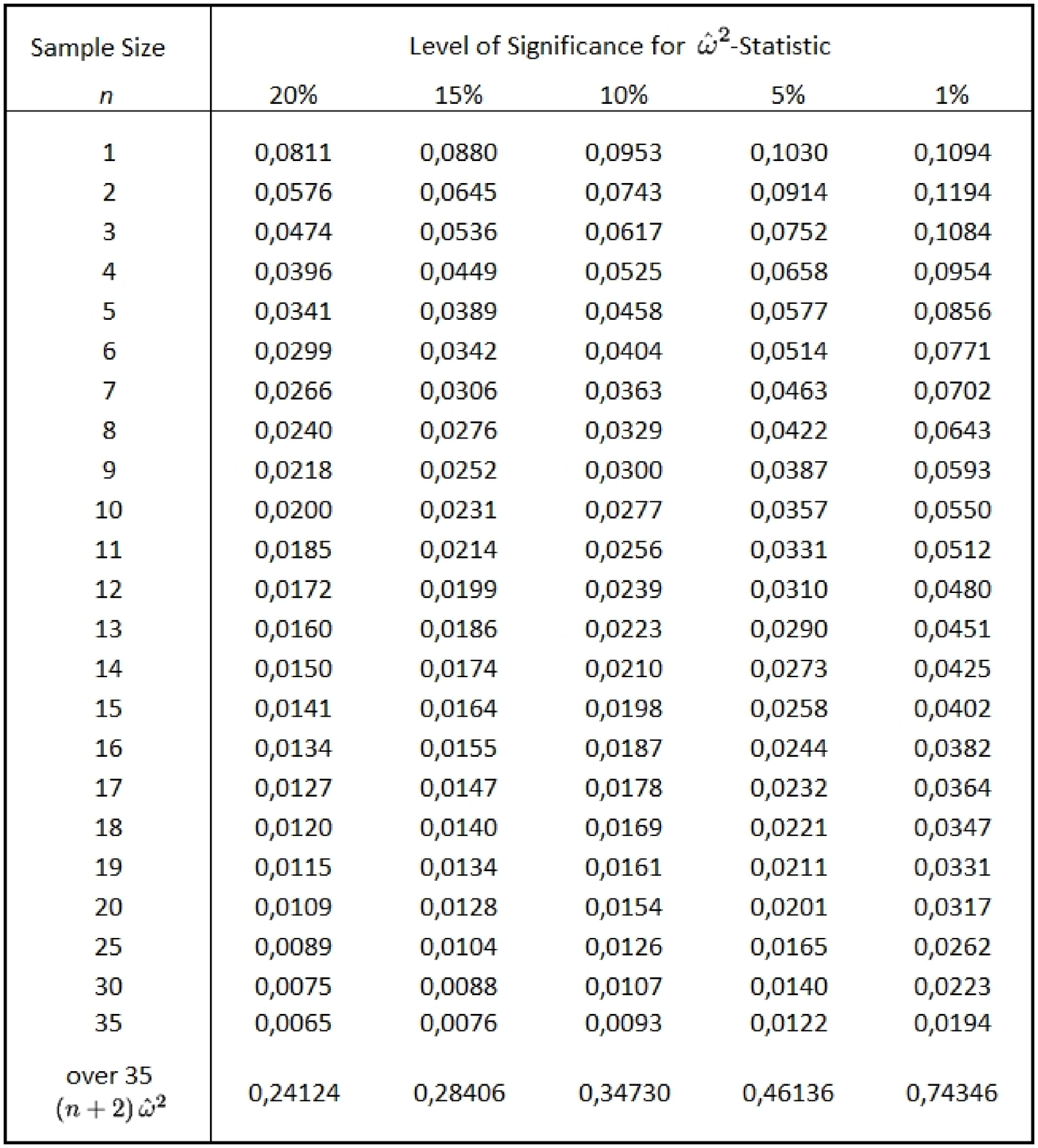}
\parbox{12cm}{
\caption{Table of critical values for the test statistic
$\hat\omega^2$ provided in Proposition 1. The quantiles are
defined by ${P(\hat\omega^2>Q_\alpha)=\alpha}$. The asymptotic values
for $n>35$ are given by the percentage points in the last line of
the table for every confidence level. The latter are to be compared
with $(n+2)\,\hat\omega^2$. The table is computed by Monte Carlo
simulations of $5\times10^7$ respective samples.} \label{percentiles}}
\end{figure*}

The critical
values of (\ref{SH}) are computed by numerical simulation and are provided in the
table of Figure\,\ref{percentiles}. For $n=1$, the critical values are $1/3$ of
the critical values for the traditional statistic. Moreover, for the 1 percent confidence level the critical values are
not strictly monotonic decreasing for increasing sample size. We checked that fact
by analytical evaluation for $n=1$ and 2. All quantiles of the table are
systematically smaller than for the quantiles of (\ref{omega2}). However, that does not necessarily
imply that there is an improvement of the type I error against the
Cram\'{e}r-von Mises statistic because for finite sample size $n$
the quantiles of both statistics have different support and the scales are not comparable.
Therefore, we compared their statistical power for a few representative examples of
distributions $F(x)$, which are supposed to be continuous and completely specified.

\begin{figure*}[htb]
\includegraphics[width=12.0cm]{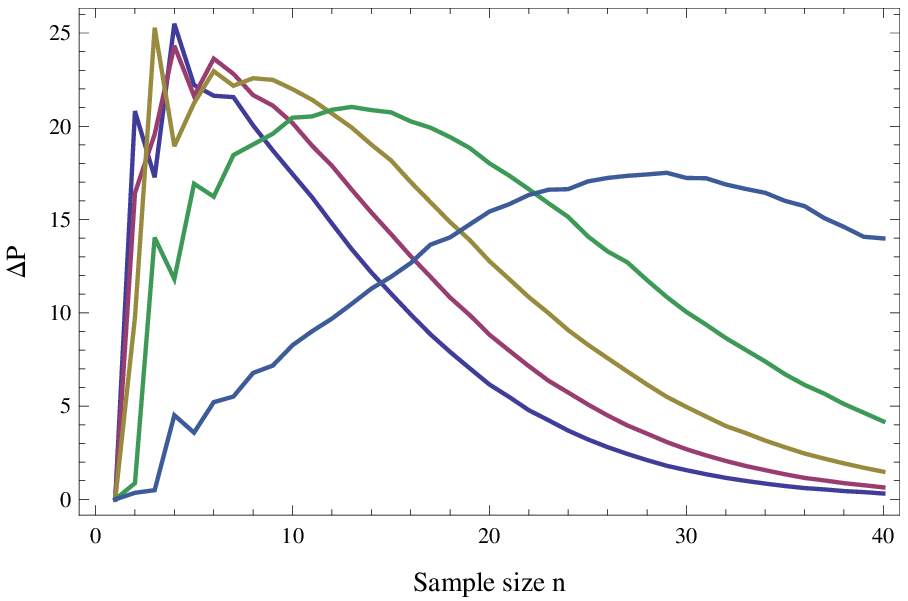}
\parbox{12cm}{
\caption{Difference $\Delta P$ for the power of the new statistic (\ref{SH}) and the traditional Cram\'{e}r-von Mises statistic (\ref{omega2}). The test is for the normal distribution ($H_0$) against uniform distribution ($H_1$). Confidence levels are 20, 15, 10, 5, 1 percentage points (from left to right). The Monte Carlo simulation is based on $10^7$ samples. The erratic behavior for very small samples is of systematical nature and not caused by insufficient sampling.} \label{fig1}}
\end{figure*}

First, we considered the test for normality ($H_0$), when $f(x)$
is the uniform distribution ($H_1$). In Figure\,\ref{fig1}, we see the
difference $\Delta P$ of the power of (\ref{SH}) and
the power of (\ref{omega2}), for confidence
levels $20, 15, 10, 5$ and $1$ percentage points. The numerical simulation has been performed
for $10^7$  Monte Carlo steps. For very small and very large
samples $n$, the power of both statistics becomes more and more equal. Intermediately, the
difference of the power between both grows until 25 percent. For all sample sizes and confidence levels
there is a higher power of (\ref{SH}) than for the
Cram\'{e}r and von Mises statistic.

\begin{figure*}[htb]
\includegraphics[width=12.0cm]{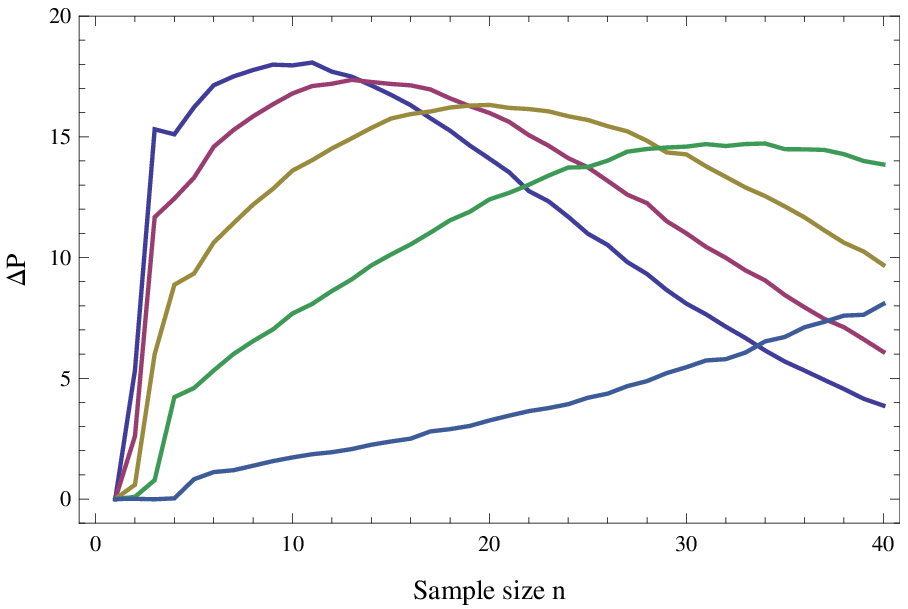}
\parbox{12cm}{
\caption{The same as in Figure \ref{fig1}, except that the test is for uniformity (see text) corresponding to Case C of \cite{St74}. The confidence levels are 20, 15, 10, 5, 1 percentage points (from left to right). The Monte Carlo simulation is based on $10^7$ samples.} \label{fig2}}
\end{figure*}

Moreover, we considered the test for uniformity given in Table 3 of Stephens study \cite{St74}.
When $F(x)$ is completely specified then $F_i$ should be
uniformly distributed between 0 and 1. The power study has therefore
been confined to a test of this hypothesis when $F_i$ is drawn from alternative distributions.
If the variance of the hypothesized $F(x)$ is correct but the mean is wrong, the $F_i$ points
will tend to move toward 0 of 1; if the mean is correct but the variance is wrong, the points
will move to each end, or will move to 1/2. Accordingly, in the Stephens study \cite{St74}, three
variants A, B and C have been defined. Case A gives random points closer to
zero than expected on the hypothesis of uniformity ($H_0$); B gives
points near 1/2; and C gives two clusters close to the boundary 0 and 1.
First, we verified the table of powers in \cite{St74} for the Cram\'{e}r-von Mises
statistic and subsequently computed the power of (\ref{SH}) for ${n=1,...,40}$.
For case A, we found that both have very similar power.
In case B, the Cram\'{e}r-von Mises statistic is slightly improved
by (\ref{SH}). Actually for case C, there is a significant improvement
for all sample sizes and all levels of confidence. The latter is shown in
Figure\,\ref{fig2}. Here, we have qualitatively the same picture as in Figure\,\ref{fig1}.
The difference of the power of both statistics is positive and reaches up to 18 percentage points.

It should be mentioned here that for case C, in the Stephens study the
Anderson-Darling statistic is superior compared to the traditional Cram\'{e}r and von Mises statistic.
If we compare the new statistic (\ref{SH}) with the Anderson-Darling
statistic for the same case C, then we find that the power of the former is
significantly higher. This might encourage further investigations of (\ref{SH}).


\end{document}